\documentclass[]{article}
\usepackage{color}
\usepackage{amsmath,amsfonts,amssymb}
\usepackage{graphicx}
\usepackage{jabbrv}
\usepackage{authblk}
\usepackage{url}
\usepackage{textcomp}
\usepackage{caption}
\usepackage{appendix}
\usepackage{hyperref}
\usepackage{gensymb}
\hypersetup{hidelinks}
\usepackage{geometry}
\usepackage{xcolor}

\usepackage{subcaption}
\usepackage{lineno}
\usepackage{gensymb}
\usepackage{marvosym}
\usepackage{makecell}
\title{Widely-tunable and narrow-linewidth hybrid-integrated diode laser at 637~nm}

\author[1,2, *]{Lisa V. Winkler}
\author[1]{Kirsten Gerritsma}
\author[1, $\dag$]{Albert van Rees}
\author[3]{Philip P. J. Schrinner}
\author[3]{Marcel Hoekman}
\author[3]{Ronald Dekker}
\author[4]{Adriano R. do Nascimento Jr.}
\author[1]{Peter~J.~M.~van~der~Slot}
\author[2]{Christian~Nölleke}
\author[1]{Klaus-J.~Boller}

\affil[1]{Laser Physics and Nonlinear Optics, Faculty of Science and Technology, MESA+ Institute, University of Twente, P.O. Box 217, 7500 AE Enschede, the Netherlands}
\affil[2]{TOPTICA Photonics AG, Lochhamer Schlag 19, 82166 Gräfelfing, Germany}
\affil[3]{LioniX International BV, Hengelosestraat 500, 7521AN, Enschede, The Netherlands}
\affil[4]{PHIX B.V., Enschede, P.O. Box 3504, 7500 DM Enschede, the Netherlands}
\affil[*]{Corresponding author: l.v.winkler@utwente.nl}
\affil[$\dag$]{currently at Chilas B.V.}

\begin{document}

\maketitle

\begin{abstract} 
We present hybrid-integrated extended cavity diode lasers tunable around 637~nm, with a gain-wide spectral coverage of 8~nm. This tuning range allows addressing the zero-phonon line of nitrogen vacancy centers and includes the wavelength of HeNe lasers (633 nm).
The lasers provide wide mode-hop free tuning up to 97~GHz and a narrow intrinsic linewidth down to 10~kHz.
The maximum output power is 2.5~mW in a single-mode fiber, corresponding to an on-chip power of 4.0~mW.
Full integration and packaging in a standard housing with fiber pigtails provide high intrinsic stability and will enable integration into complex optical systems.
\end{abstract}

\section{Introduction}

Photonic quantum technologies, such as photonic quantum processing \cite{slussarenko2019photonic}, quantum-key distribution \cite{sibson2017chip} and quantum sensing\cite{degen2017quantum}, are currently undergoing a transition from research labs to industrial applications \cite{moody20222022}. Upscaling of such systems \cite{Taballione2023modeuniversal, somhorst2021quantum} calls for fully integrated on-chip laser sources \cite{mahmudlu2023fully}. While integrated lasers are already fairly mature in the infra-red, many applications require lasers in the visible range, e.g., for addressing particular atomic and ionic transitions \cite{pogorelov2021compact}, quantum dots or nitrogen vacancy (NV) centers \cite{shields2015efficient}.

Chip integration of suitable lasers can be achieved by exploiting the wide transparency range and high index contrast of the silicon nitride platform, which provides low-loss waveguide circuits \cite{roeloffzen2018low} both in the infrared and in the visible range. The first realization of a hybrid integrated laser in the visible \cite{franken2021hybrid} has recently been followed by extending self-injection locking (SIL) \cite{jin2021hertz} to several visible wavelengths \cite{corato2022widely, siddharth2022near}.
Although of fundamental interest, the latter three demonstrations relied on separate chips aligned with manual stages for edge coupling, which limits wavelength and power stability. We note that hybrid integration followed by packaging are key steps to let the laser performance live up to the promise of high intrinsic stability associated with photonic integrated circuits. These steps are also instrumental for active stabilization \cite{lockingAlbertLisa}, portability and further integration into more complex optical systems.

For gain-wide tunability, a large mode-hop free tuning range and a narrow-linewidth, a suitable laser concept needs to be chosen. Both, extended cavity diode lasers (ECDLs) \cite{fan2020hybrid} and lasers based on SIL \cite{jin2021hertz} can offer narrow intrinsic linewidths. However, SIL lasers rely on narrowing a pre-defined mode of the laser diode, which limits their continuous tuning range \cite{kondratiev2017self}, with typical values of a few GHz \cite{guo2022chip, snigirev2023ultrafast}. ECDLs on the other hand are free from this restriction and solely rely on tuning of the extended cavity. Therefore, they can provide wide mode-hop free tuning \cite{vanRees:20} which can span the full gain bandwidth of the amplifier \cite{wandt1996external}.

Here we present a visible-range, fully integrated ECDL with permanently bonded gain and feedback waveguide circuits. For reliable operation and portability, the integrated laser is electronically, thermally, and photonically packed in a standard housing using wirebonding, a thermo-electric element and output fibers. The laser consists of a semiconductor amplifier that receives optical feedback from a Si\textsubscript{3}N\textsubscript{4} waveguide circuit, acting as a frequency-selective mirror. The circuit parameters are selected to provide tunability around a center wavelength of 637 nm, for addressing NV centers in diamond. Regarding the selection of the feedback circuit, we note that tuning to 633 nm is also of interest for classical metrology, such as to replace HeNe lasers.

\section{Design}

\subsection*{Waveguide cross-section}
For constructing narrow-linewidth lasers with gain-wide tunability and a high side-mode suppression, choosing a suitable waveguide cross-section is crucial. For a narrow laser linewidth, low propagation losses are key as they enable a long photon lifetime in the cavity. For single-frequency oscillation with high side-mode suppression, and for expanding tunability across the full gain bandwidth of the diode amplifier, small micro-ring resonators are required. Therefore the chosen waveguide cross-section must allow low bend radii while maintaining a low propagation loss. For efficient coupling to the gain section and to output fibers, the chosen cross-section also has to facilitate matching the corresponding mode fields using tapering.
While silicon nitride waveguides for the red wavelength range have been demonstrated \cite{sinclair2020silicon, chauhan2021visible, corato2022widely}, exhibiting different choices for the trade-off between propagation loss and bend loss, we here optimize the cross-section specifically for the described goals.

\begin{figure}[tbh]
     \centering
     \begin{subfigure}[b]{0.45\textwidth}
         \centering
         \includegraphics[width=\textwidth]{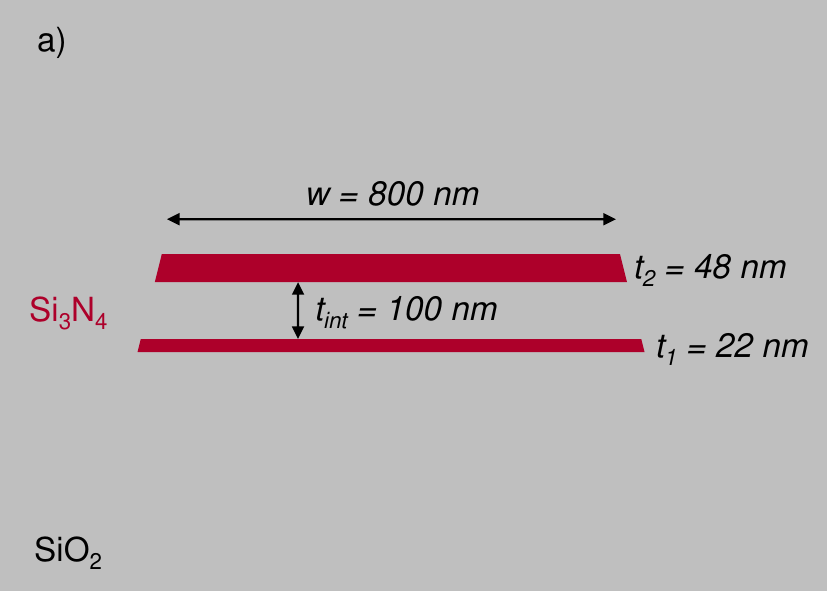}
         \label{fig:cross_Section}
     \end{subfigure}
     \hfill
     \begin{subfigure}[b]{0.45\textwidth}
         \centering
         \includegraphics[width=\textwidth]{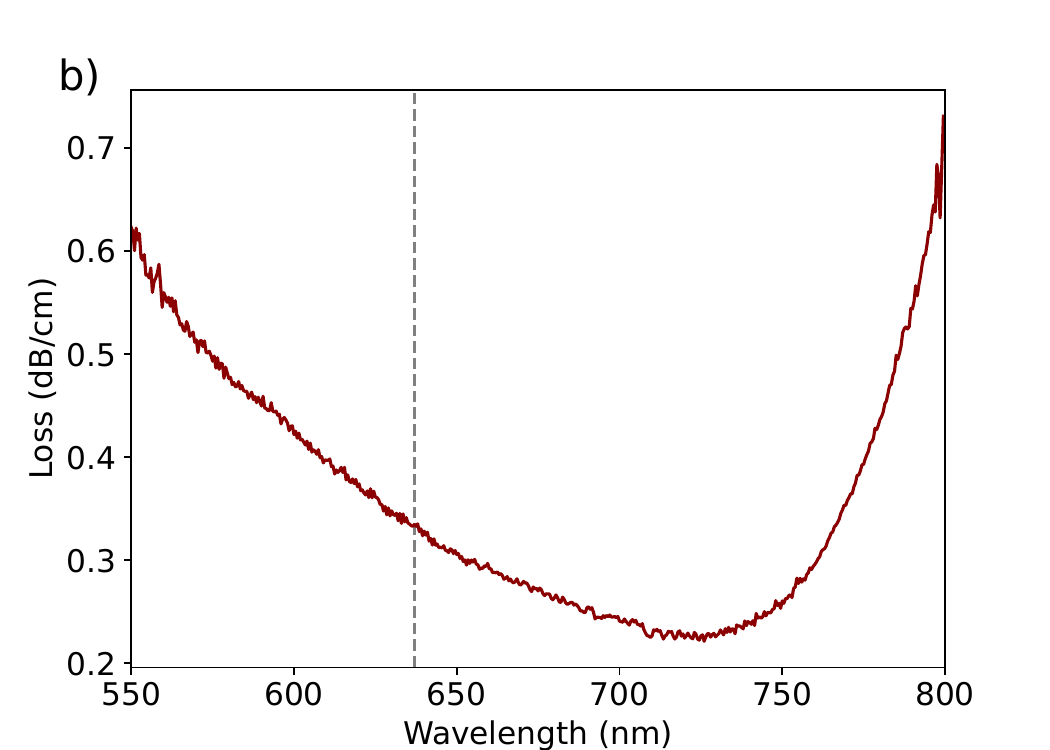}
         \label{fig:spiral_loss}
     \end{subfigure}
     \hfill
        \caption{(a) Double-stripe waveguide cross-section. The layerstack is optimized for low propagation loss while maintaining a negligible bend loss for bend radii above 100~$\mu$m, single-mode operation and allowing efficient coupling to the gain section and fibers. (b) Propagation loss measured using a 47-cm spiral. At the target wavelength of 637~nm, the propagation loss is 0.33 dB/cm (vertical dashed line).}
        \label{fig:cs-loss}
\end{figure}

It is well known that the main contribution to propagation losses will stem from sidewall scattering \cite{Bauters:11} as the waveguide's sidewalls are much rougher due to the etching process than the smooth top and bottom interfaces. We estimated losses from sidewall scattering via path integrals of the mode intensity along the waveguide's sidewalls and calibrated the estimation using loss data from previous fabrication runs. To avoid scattering into higher-order modes, the design was restricted to cross-sections that guide only a single transverse mode. Mode field distributions were calculated with a 2D finite difference mode solver (OptoDesigner) to optimize the core geometry parameters along the following guidelines: Increasing the aspect ratio by thinning the Si\textsubscript{3}N\textsubscript{4} core minimizes loss from sidewall scattering \cite{bogaerts2012silicon}. However, thicker cores and a double-stripe geometry (compared to a single stripe) provide tighter guiding for minimizing bend loss. We restricted the optimization to double-stripe cross sections with negligible bending loss ($\le$ 0.01~dB/cm) for bend radii down to 100 $\mu$m. Spot size converters were designed, using lateral tapering for mode matching to a semiconductor optical amplifier (SOA) and a combination of lateral and vertical tapering for mode matching to output fibers. Fig. \ref{fig:cs-loss}(a) shows the optimized cross section comprising two Si\textsubscript{3}N\textsubscript{4} cores with thicknesses of 22 and 48 nm, with a 100-nm spacing and a waveguide width of 800~nm.

To provide high fabrication fidelity at a small feature size, and to obtain maximally smooth sidewall surfaces, fabrication was carried out using DUV stepper lithography. To verify whether the intended properties of the fabricated cross section and circuit components match the target values, an experimental characterization is performed before integration with an SOA. Of central interest is quantifying the propagation loss, as it is crucial for the laser threshold and output power and can limit the minimum achievable laser linewidth \cite{photonics7010004}. The propagation loss was determined by measuring the transmission through a 47-cm long waveguide spiral with a minimium bend radius of 100 $\mu$m in comparison to a short straight waveguide using a supercontinuum source. Fig. \ref{fig:cs-loss}(b) shows the measured propagation loss vs. wavelength, yielding a value of 0.3$\pm$0.1~dB/cm at the target wavelength of 637~nm. This corresponds to round-trip propagation losses of 0.9 and 3.8~dB, respectively for the lasers described below and a minimum achievable intrinsic linewidth of approximately 1~kHz \cite{photonics7010004}. The lowest loss occurs at around 725 nm (0.23~dB/cm). With increasing wavelength, the propagation loss first decreases due to decreasing confinement. From wavelengths above 720~nm, the measured loss increases as the decreasing confinement causes the bend loss to dominate. From this measurement, we conclude that bend loss is indeed negligible as intended at our target wavelength for radii down to 100~$\mu$m, with a margin of error included to allow for fabrication variation.

\subsection*{Laser design}

\begin{figure}[ht]
\centering\includegraphics[width=10cm]{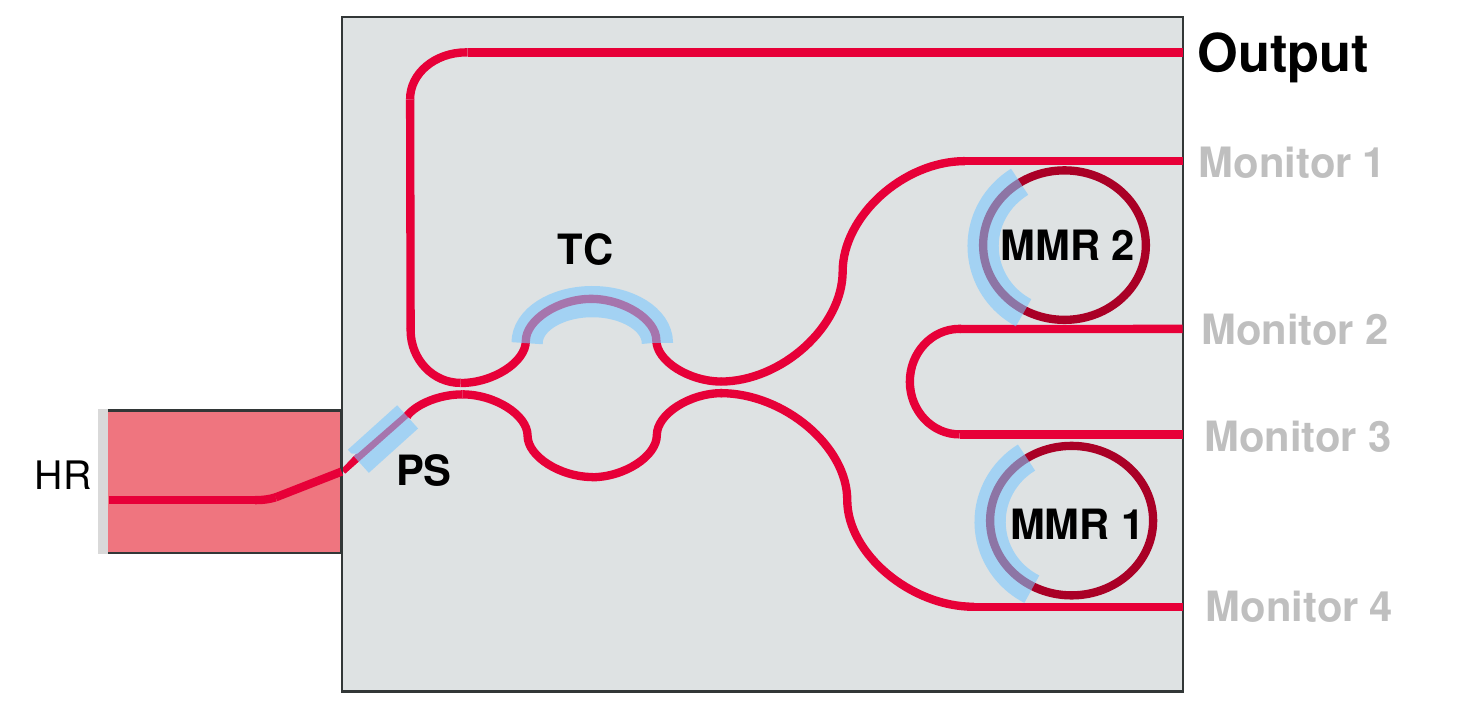}
\caption{Schematic drawing of the laser circuit. Red: Semiconductor optical amplifier. Grey: Si\textsubscript{3}N\textsubscript{4}/SiO\textsubscript{2} feedback chip with phase section (PS), tunable coupler (TC) and microring resonators MMR1 and MMR2. Resistive heaters for thermal tuning are shown in blue.}
\label{fig:schematic}
\end{figure}

The lasers consist of a gallium arsenide semiconductor optical gain chip (EXALOS, custom device with a peak wavelength of 638~nm) which is hybrid-integrated with a Si\textsubscript{3}N\textsubscript{4}/SiO\textsubscript{2} feedback chip to form an ECDL. Fig. \ref{fig:schematic} shows a schematic drawing of the laser circuit.

The back facet of the amplifier is coated with a highly reflective (HR) coating ($>$95\%), forming one mirror of the laser cavity.
The second cavity mirror is a loop mirror, formed by two micro-ring resonators (MRRs) on the feedback chip. These MRRs with respective radii of 100 and 103~$\mu$m also act as a Vernier filter, providing frequency-selective feedback. By inducing multiple optical roundtrips in the MRRs, they also extend the resonator photon lifetime to lower the laser linewidth. While the individual MRRs have a respective free spectral range (FSR) of only approx. 0.35~nm each, they combine to form a Vernier filter with a wide FSR of 14~nm.
This wide Vernier FSR enables selecting a single wavelength for laser operation within the approximately 9-nm wide gain spectrum of the diode amplifier.
A tunable coupler (TC) placed in the cavity allows coupling light out of the cavity to the output port and provides adjustable feedback to the gain section. By coupling out in this manner, only light that has passed the Vernier filter is transmitted to the output port, which reduces the level of unwanted amplified spontaneous emission (ASE).
The output port, as well as the through and drop ports of the MRRs (monitor ports) are coupled to polarization-maintaining single-mode fibers (PM630-HP, FC/APC) in a fiber array. To prevent back reflections into the laser cavity, the SiN chip and fiber array facets were polished at an angle of 8 degrees.
Resistive heaters are placed on top of the feedback chip for thermal tuning of the individual elements. Specifically, the transmission of the Vernier filter can be tuned by adjusting the heaters on the ring resonators, while a phase section (PS) heater can be adjusted to align the cavity mode with the Vernier mode \cite{vanRees:20}.

The power coupling coefficient $\kappa^2$ from the bus waveguides to the MRRs is an important design choice as it determines the trade-off between a narrow linewidth on the one hand and higher stability and tunability on the other hand: Choosing a low coupling strength decreases the laser linewidth by increasing the cavity length (and thus the photon lifetime) with a higher number of roundtrips in the MRRs. On the other hand, too low a coupling strength decreases the amount of feedback to the gain section by increasing the losses in the feedback circuit. The result is a larger susceptibility to undesired broadband feedback, such as from spurious scattering or facet reflections. Furthermore, resonance splitting and back reflections from the MRRs increase for a lower $\kappa^2$ \cite{li2016backscattering}.

For a comparison of these effects, we explore two variants with largely different power coupling to the ring resonators ($\kappa^2 = 1.6$\% and $\kappa^2 = 13$\%). Due to the different number of roundtrips in the MRRs, the lasers strongly differ in effective optical round-trip length, with approximately 26 and 114~mm for the high-coupling and the low-coupling laser, respectively. A narrower linewidth is expected for the longer laser.

\begin{figure}
     \centering
     \begin{subfigure}[b]{0.47\textwidth}
         \centering
         \includegraphics[width=\textwidth]{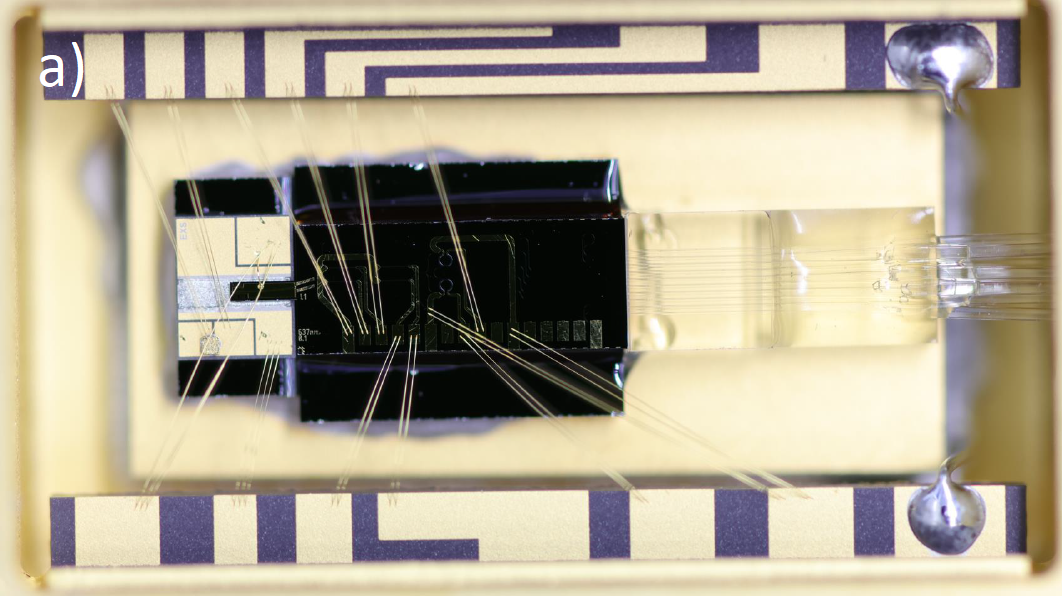}
         \label{fig:assembly_overview}
         
     \end{subfigure}
     \hfill
     \begin{subfigure}[b]{0.47\textwidth}
         \centering
         \includegraphics[width=\textwidth]{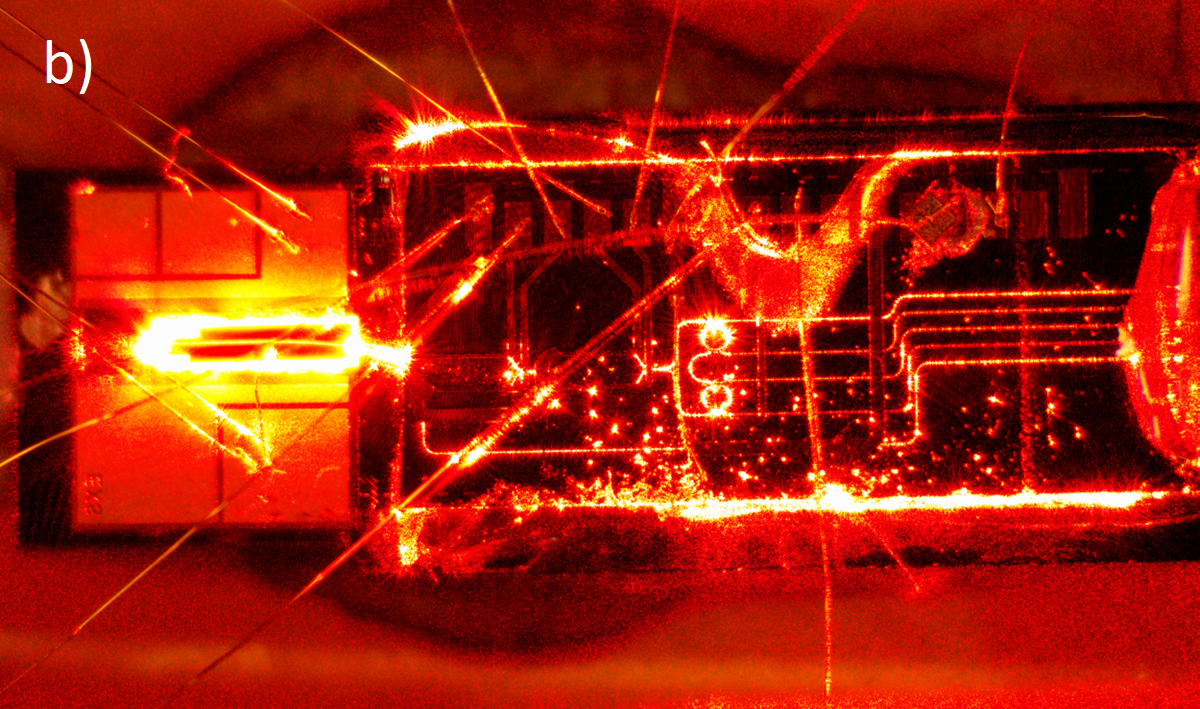}
         \label{fig:laser_photo_on}
     \end{subfigure}
     \hfill
        \caption{(a) The packaged high-coupling laser in a standard 14-pin butterfly housing (b) Close-up of the low-coupling laser in operation. Left: Gain chip on submount. Right: Feedback chip with both ring resonators lighting up.}
        \label{fig:laser-photos}
\end{figure}

The gain section and the Si\textsubscript{3}N\textsubscript{4}/SiO\textsubscript{2} feedback chip are hybrid integrated and packaged. For hybrid tunable lasers in the visible range, the photonic packaging crucially depends on identifying compatible bonding materials (e.g. epoxies) to ensure efficient light transmission. At the same time, thermal expansion coefficients and bonding techniques must be carefully considered to guarantee a reliable assembly and long-term stability. In this work, the device was packaged in a standard 14-pin butterfly package with glass lid, a thermoelectric cooler (TEC), NTC thermistor, and a 9-channel fiber array to monitor all ports, including alignment loops. Fig. \ref{fig:laser-photos}(a) shows the full laser module, while \ref{fig:laser-photos}(b) shows a close-up of the laser in operation.

\section{Laser characterization}
We characterize the assembled lasers with respect to output power, tunability and linewidth. For a convenient comparison between the high-coupling and the low-coupling laser, all results are summarized in Table \ref{tab:summary}.

\subsection*{Output power}

The output power of both lasers is measured as a function of the pump current. The Vernier filter is tuned for maximum laser output power, to the respective wavelengths of 639 and 638~nm.
To confirm that the lasers stay at these wavelengths for all current settings, their spectrum is monitored using an optical spectrum analyser (OSA, ANDO AQ6315A). For each pump current, the phase section is fine tuned for maximum output power. For the high-coupling laser, the output power is measured in the output port.
For the low-coupling laser however, the highest output was found in the monitor  port 1, so the output power in this port was also measured.
For the low-coupling laser, the output power is recorded up to the maximum specified current of 70~mA. For the high-coupling laser, a gain chip with a lower specified maximum current was used, so we limited the output current to 50~mA. All powers are measured using a power sensor (Thorlabs S120C) and a power meter console (Thorlabs, PM100D).
To convert the fiber-coupled output powers into on-chip powers, the coupling loss was measured using on-chip alignment loops which are connected to the output fiber array.

\begin{figure}[tbh]
    \centering
    \includegraphics[width=7.5cm]{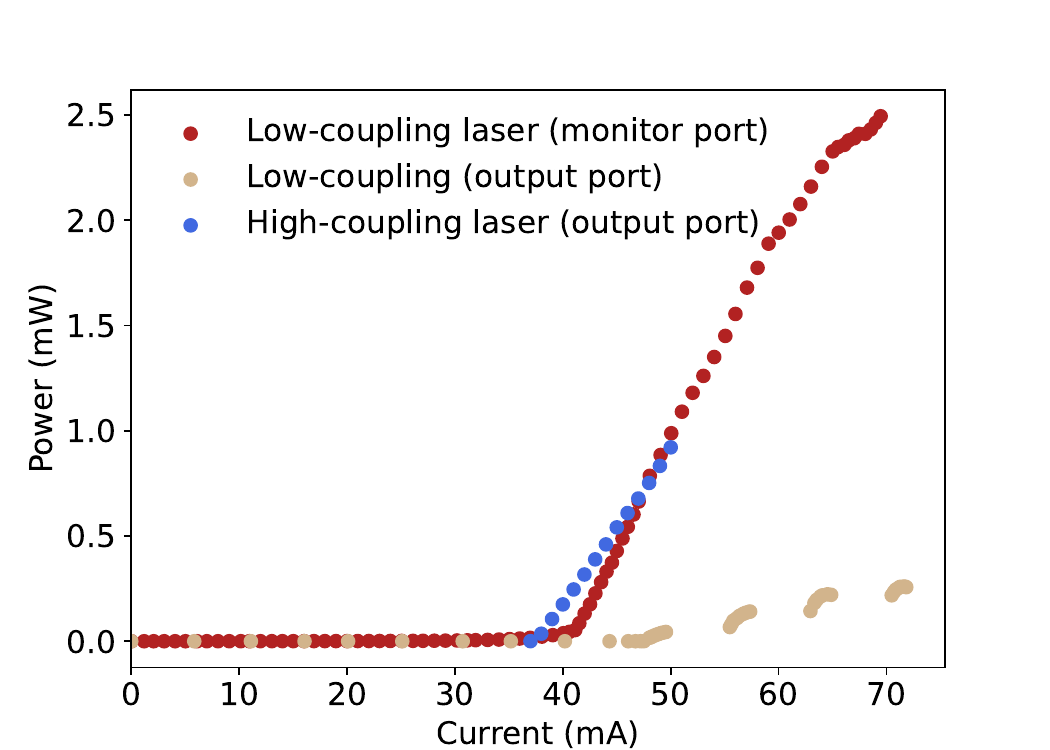}
    \caption{Fiber-coupled output power of both lasers with a measurement uncertainty of $\pm 5\%$. The high-coupling laser has a maximum output power of 0.92~mW while the low-coupling laser has a maximum power of 2.5~mW in monitor port~1 and a power of 0.25~mW in the output port.}

    \label{figure:PI}
\end{figure}

Fig.~\ref{figure:PI} shows the measured output power of both lasers as a function of the pump current. Above threshold, the power rises approximately linearly, as expected.
A linear fit yields thresholds of 37.5 and 40.7~mA and slope efficiencies of 0.07 and 0.10~W/A for the high-coupling and the low-coupling laser, respectively.
For the low-coupling laser, the output power reaches a maximum output power of 2.5~mW in monitor port~1.  Taking the coupling loss of 2.1$\pm$0.3~dB into account, this corresponds to an on-chip power of 4.0~mW.  We attribute the lower power of 0.25~mW in the output port to undercoupling of the MRRs, which reduces resonant power transmission through the MRRs \cite{yariv2002critical}. We therefore decide to use light from monitor port~1 for all further characterization measurements of this laser.
For the high-coupling laser, the measured maximum output power was 0.92 mW, which is approximately 1.3~mW on-chip, using an estimated coupling loss of 1.6$\pm$0.3 dB.

\subsection*{Frequency noise and linewidth}
The frequency noise of the lasers was measured using a delayed self-heterodyne setup with a 20-m fiber delay and a 40-GHz acousto-optic modulator (AA Opto-Electronic MTS40-A3-750.850).
The beat signal was recorded using an oscilloscope (Teledyne Lecroy HD06104A) and converted into a single-sided noise frequency noise power spectral density \cite{Schiemangk:14, Tsuchida:17}.
The noise spectra of both lasers were measured near the respective gain maximum.
To decrease the intrinsic laser linewidth \cite{fan2020hybrid}, the measurement was performed at the maximum currents deemed safe for the respective gain chips (70 mA and 55 mA).

\begin{figure}[tbh]
\centering
\includegraphics[width=7.5cm]{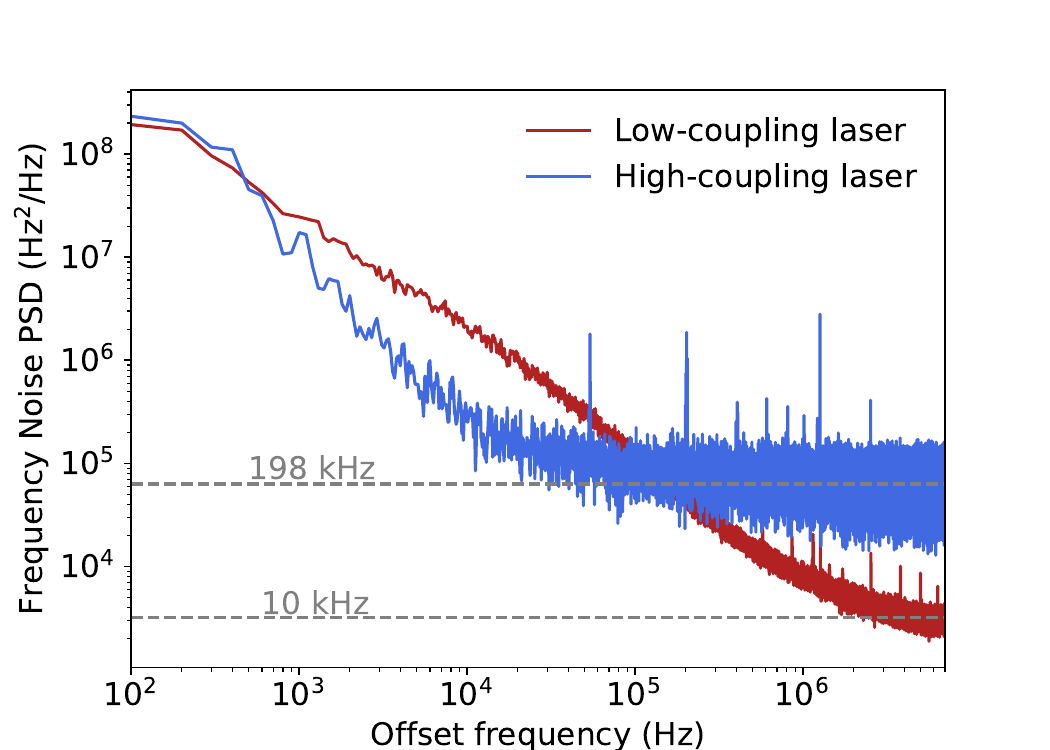}
\caption{Measured power spectral density of frequency noise.
The frequency noise of the high-coupling laser levels off at a white noise level of 63~kHz$^2$/Hz. This noise level corresponds to an intrinsic laser linewidth of 198~kHz, obtained by multiplying with $\pi$. For the low-coupling laser, the lowest measured noise level is 3~kHz$^2$/Hz, corresponding to an intrinsic linewidth below 10~kHz.
}
  \label{fig:linewidth}
\end{figure}

The frequency noise power spectral density for both lasers is shown in Fig. \ref{fig:linewidth}. 
It can be seen that their noise is decreasing with increasing offset frequency. 
The spectrum of the high-coupling laser levels off at lower offset frequencies (around 100~kHz) towards a flat, white-noise spectrum.
The decrease of the noise density is typical for technical noise, while the white noise level is associated with the intrinsic linewidth of the laser, also called the Schawlow-Townes linewidth, or fast linewidth component \cite{schawlow1958infrared}.
The white noise level of the high-coupling laser corresponds to an intrinsic laser linewidth of 198$\pm$50~kHz.
For the high Q laser, the noise spectrum reaches about 20-times lower values and full leveling-off seems to occur beyond the upper edge of the detection bandwidth.
The lowest noise level thus provides an upper limit for the laser's white noise. This corresponds to an intrinsic linewidth below 10~kHz as indicated by the lower dashed line in Fig. \ref{fig:linewidth}.
We attribute the lower intrinsic linewidth of the low-coupling laser mainly to the longer photon lifetime in the cavity. Due to the lower power coupling to the MRRs, the cavity length is a factor of 4 higher, leading to an expected linewidth reduction by a factor of about 16 \cite{schawlow1958infrared}.
A second reason for a lower linewidth can be found in the higher output power which we estimate to contribute another factor of two.

\subsection{Wavelength Coverage}
Tuning a laser to different target wavelengths is crucial for many applications, for example for addressing specific transition frequencies of point defects in crystal structures, which can shift, depending on their environment. Furthermore, a wide wavelength coverage can make one laser suitable for multiple applications instead of requiring a specific device for each individual target wavelength.

To investigate the tuning range of the lasers, their output spectra were recorded for different heater settings using the optical spectrum analyzer. The spectra were obtained by coarse tuning one of the ring resonators in steps of approximately 11~mW applied heater power, causing mode hops corresponding to the free spectral range of 0.35~nm. After each hop, the output power was maximized by fine tuning the ring resonators and the phase section for maximum output power. We note that the low-coupling laser occasionally shows multi-mode operation based on backscattering of a single MRR as described in \cite{li2016backscattering}. With a FWHM of 2~GHz, the transmission peaks of the MRRs in the low-coupling laser are also narrower, compared to 14~GHz in the high-coupling laser. Accordingly, more accurate tuning of the heater elements is required to achieve tuning over the whole gain bandwidth for the low-coupling laser.

\begin{figure}[tbh]
        \centering
        \begin{subfigure}[b]{0.495\textwidth}
            \centering
            \includegraphics[width=\textwidth]{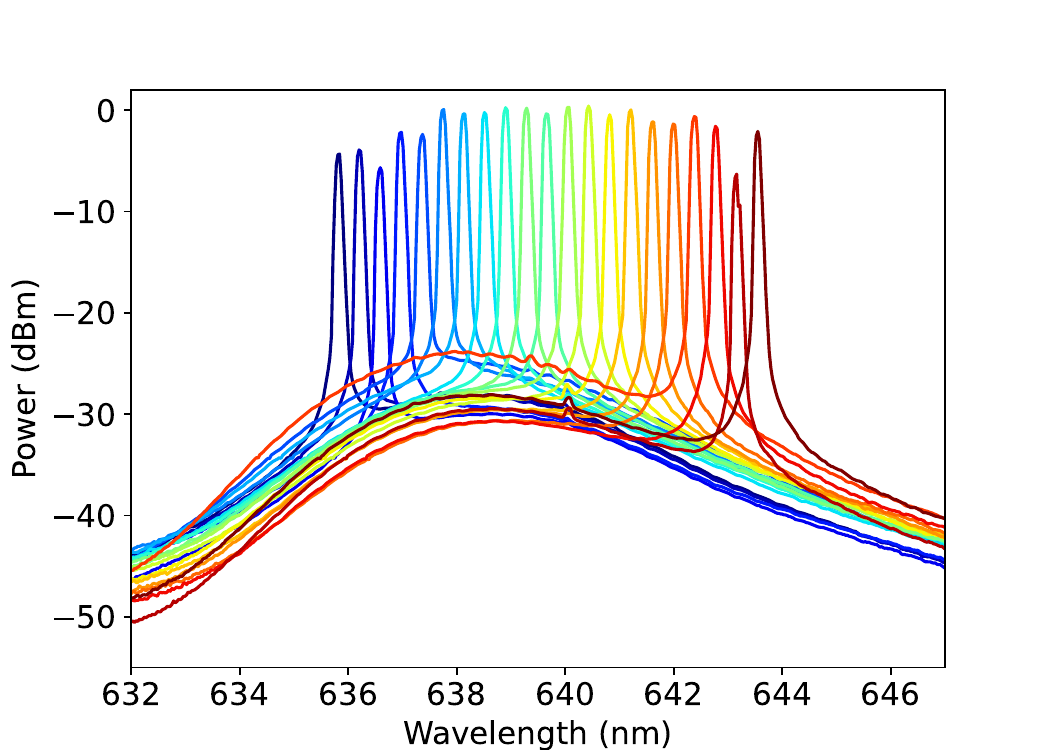}
            \caption[]%
            {{Low-coupling laser, measured in the monitor port}}    
            \label{figure:WL_A2}
        \end{subfigure}
        \hfill
        \begin{subfigure}[b]{0.495\textwidth}  
            \centering 
            \includegraphics[width=\textwidth]{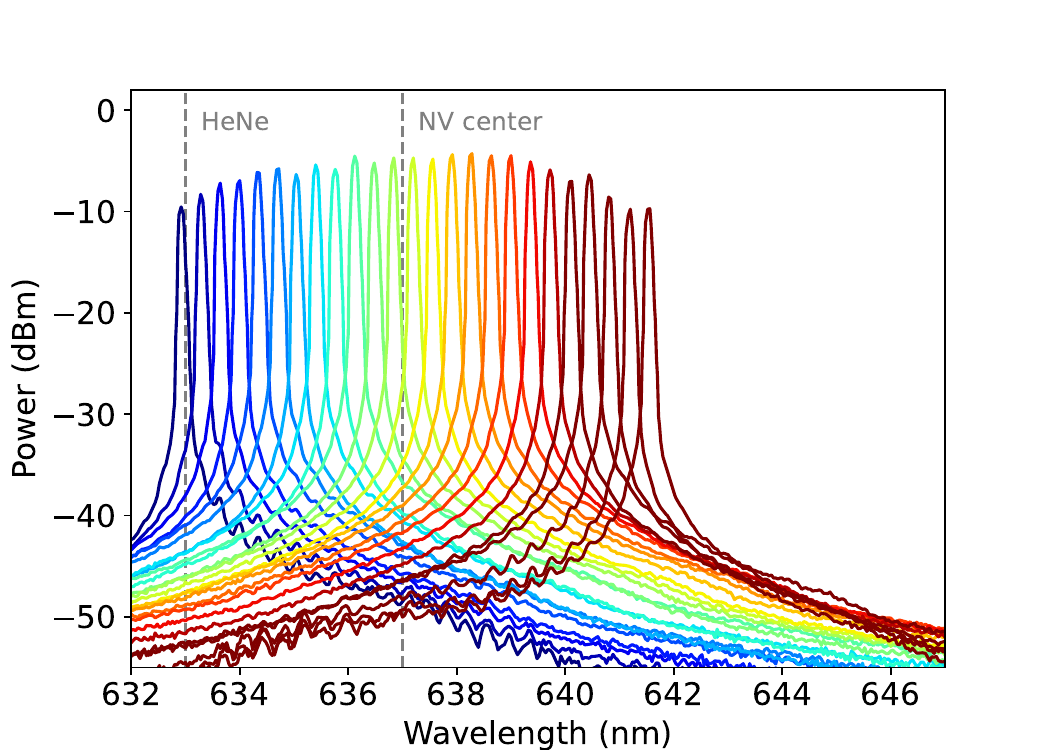}
            \caption[]%
            {{High-coupling laser, measured in the output port}}   
            \label{figure:WL_A4}
        \end{subfigure}
                \hfill
        \begin{subfigure}[b]{0.495\textwidth}  
            \centering 
            \includegraphics[width=\textwidth]{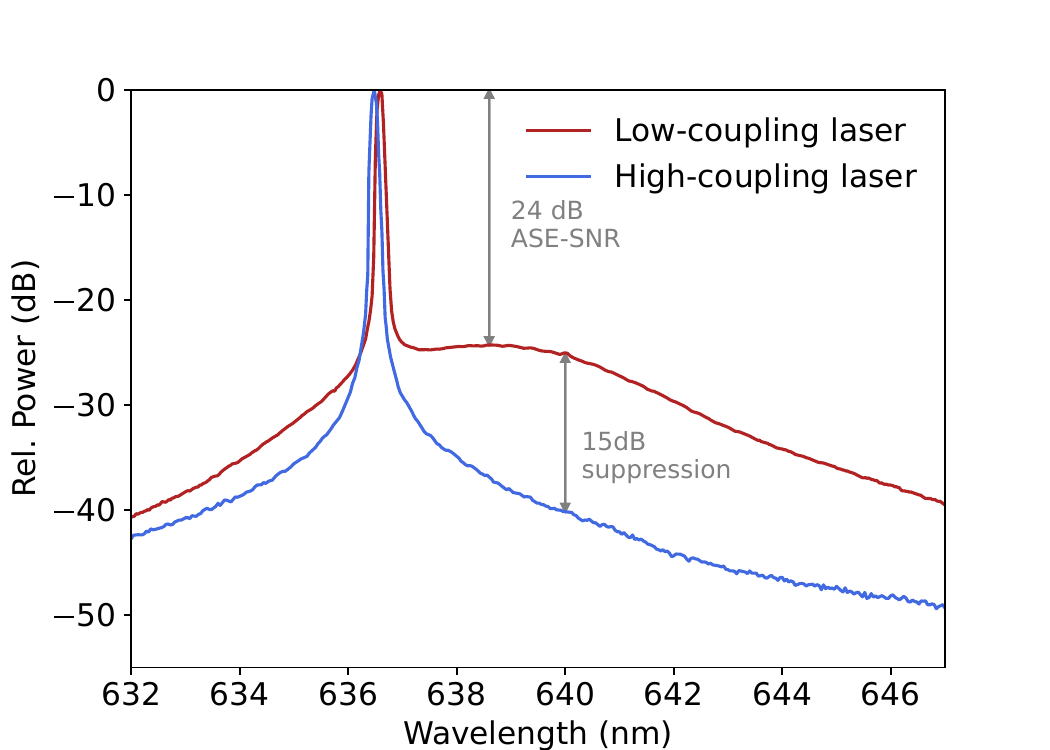}
            \caption[]%
            {{Comparison of the ASE suppression in both lasers}}   
            \label{figure:SMSR}
        \end{subfigure}
        \caption{Superimposed output spectra obtained by tuning the laser wavelength via coarse-tuning one MRR heater and then fine-tuning the MRR and PS heaters for maximum output power. The spectral resolution is 0.1 nm and the power accuracy is 0.3 dB. a) Measured from monitor port~1 of the low-coupling laser at a pump current of 68.9~mA. The laser can be tuned from 635.8 to 643.6~nm b) Measured from the output port of the high-coupling laser at a pump current of 50.0~mA. The laser can be tuned from 632.9 to 641.5~nm. c) Comparison of emission spectra from the low-coupling and the high-coupling laser, normalized to the same peak value. In the low-coupling laser, the ASE-SNR is 24~dB. In the high-coupling laser the ASE is suppressed by an additional 15~dB.}
        \label{fig:wl-coverage}
    \end{figure}

Fig.~\ref{fig:wl-coverage} shows superimposed output spectra recorded for the different heater settings.
Both lasers provide approximately the same spectral coverage of around 8~nm, only limited by the gain bandwidth of the laser diode. The low-coupling laser can be tuned over 7.7~nm with a center wavelength of 639.7~nm, while the high-coupling laser can be tuned over 8.6~nm, with a center wavelength of 637.2~nm. This enables addressing the zero-phonon line of NV centers as well as addressing the emission wavelength of Helium-Neon lasers. We attribute the approximate 3-nm redshift of coverage in the low-coupling laser to a higher temperature of the gain chip, caused by a more distant placement of the temperature sensor. This indicates that that the tuning range can be further extended by changing the temperature of the gain chip.
In addition to the slightly smaller tuning range, the low-coupling laser also shows a larger variation in output power which can be explained by the higher sensitivity to parameter changes as described above.

Besides wide laser tuning,  Figs. \ref{fig:wl-coverage} a) and b) also display some broadband background caused by ASE. For a direct comparison Fig. \ref{fig:wl-coverage} c) shows one spectrum obtained for each laser, normalized to the same peak value.
In the low-coupling laser, ASE corresponding to the gain spectrum is clearly visible, with a signal-to-ASE-noise ratio (ASE-SNR) of approximately 25~dB. The high-coupling laser, on the other hand, shows a much lower ASE contribution.
The difference can be explained by having used different ports for output coupling (see also Fig. \ref{fig:schematic}): In the low-coupling laser, light is coupled out from monitor port~1, while the output port is used for the high-coupling laser. ASE generated in the gain section is for the most part not on resonance with the MRRs. Therefore, ASE is barely transmitted through the MRRs, but instead to the monitor ports.

\subsection*{Mode-hop free tuning}
Continuous, mode-hop free tuning of the laser frequency is essential for spectroscopy applications and is also a crucial prerequisite for laser frequency stabilization. Here we present a first demonstration of mode-hop free tuning of integrated ECDLs in the visible.

Tuning is achieved by controlling three heater elements: MMR1 and MMR2, forming the Vernier filter\, coarsely set the wavelength as desired within the gain profile. The phase section PS can be used to adjust the optical cavity length, allowing for small changes in the laser frequency. By tuning a single one of these heater elements, the laser can be tuned continuously only over approximately one FSR of the laser cavity (typically a few GHz). By tuning all three heater elements \cite{vanRees:20}, a much larger tuning range can be achieved \cite{pintus2023demonstration}.

\begin{figure}
     \centering
         \includegraphics[width=7.5cm]{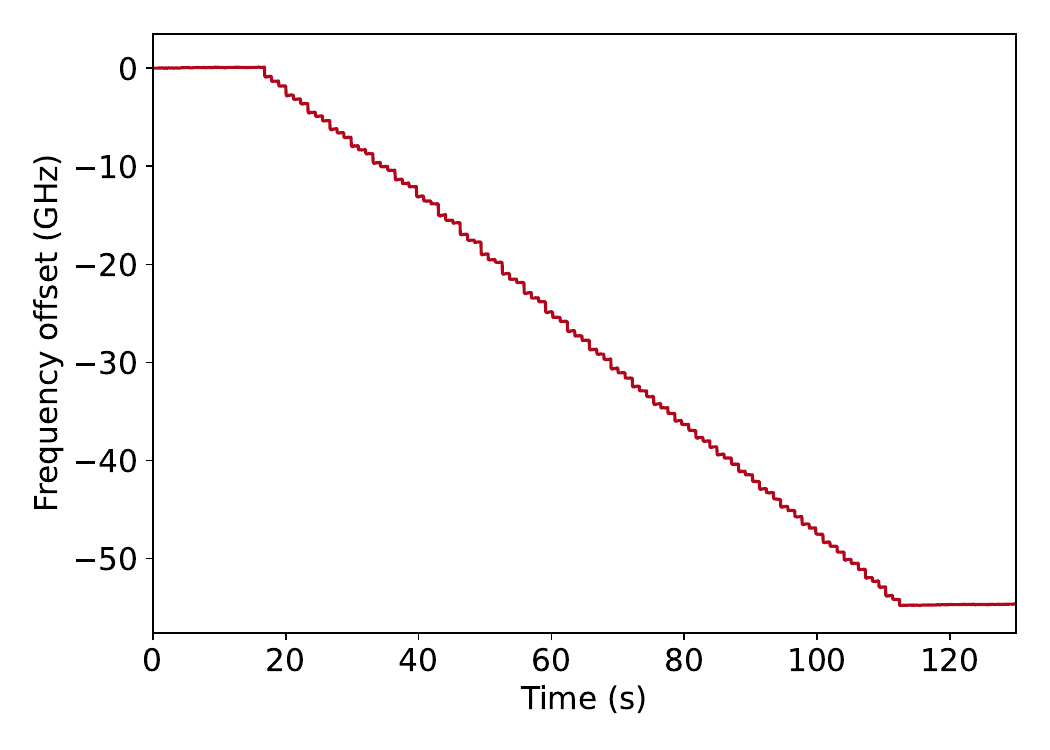}
        \caption{Automatic mode-hop free tuning of the high-coupling laser by synchronous tuning of the PS and MRRs, measured in the output port. The absolute change in the cavity frequency is presented as function of the measuring time. The wavelength shifts from 639.30~nm up to 639.37~nm, which corresponds to a mode-hop free tuning range of 55~GHz with respect to the starting wavelength.}
        \label{fig:MHFT}
\end{figure}

To investigate the capability for automated tuning, we used the high-coupling laser and changed the heater powers in a fixed ratio (1:1:8 for MRR1, MRR2 and PS, respectively). The three heater powers were ramped up linearly vs. time, using  small steps of 2.3 and 18~mW for MMRs and PS respectively, up to a maximum increase of 72~mW (MRRs) and 580~mW (PS).
The stepping of heater currents was carried out using a computer via serial commands to a USB interface.
Fig.~\ref{fig:MHFT} shows the recorded laser wavelength over time, demonstrating mode-hope free tuning over 55~GHz, with the steps corresponding to an adjustment of the heaters, followed by a 1-s break. This feed-forward automatic tuning demonstrates that the laser tunes in a predictable manner and does not need active feedback, e.g., in the form of monitoring the output power in the different monitor ports, for wide mode-hop free tuning. 

To investigate the maximum mode-hop free tuning range of the lasers, they were also tuned manually, adjusting the heater voltages while monitoring the emission spectrum and power, as well as visually checking the brightness in different parts of the circuit from the top, using a microscope. The low-coupling laser could be tuned continuously over 43~GHz, before a mode hop occurred. We address this to lower feedback to the gain section as described above.
The high-coupling laser, on the other hand, provided an even wider mode-hop free tuning range. Manual tuning using all available heater elements (MMR1, MRR2, PS and TC) allowed mode-hop free tuning over 97~GHz in total.
Overall, the tuning of this laser was limited by the cooling rate of the thermoelectric cooler which needs to dissipate the heat generated by the heater elements.

\begin{center}
\begin{tabular}{ |c||c|c|c|c|  }
\hline
\(\kappa^2\) & $\lambda$ & $\Delta\nu\textsubscript{intr.}$ & MHFT range & max. $P\textsubscript{out}$\\
\hline
\hline
13\%   & \makecell{632.9-641.5~nm\\($\Delta$ 8.6~nm)}    &198~kHz&\makecell{97~GHz (man.)\\55~GHz (auto.)} & 0.92~mW\\
\hline
1.6\% & \makecell{635.8-643.6~nm\\($\Delta$ 7.7~nm)}  & $<$10~kHz & 43~GHz (man.) & \makecell{2.5~mW}\\
\hline
\end{tabular}
\captionof{table}{Overview of the measured properties of the low-coupling and the high-coupling laser.}
\label{tab:summary}
\end{center}

\section{Conclusion and outlook}

We have demonstrated the first hybrid-integrated extended cavity diode laser, operating around a center wavelength of 637~nm. We have realized and compared two laser variants, with different power coupling coefficients to the micro-ring resonators: With a lower coupling coefficient to the MRRs, we obtained an intrinsic linewidth below 10~kHz, wide tuning over 7.7~nm and a mode-hop free tuning range of 43~GHz. With a higher coupling coefficient, the tunability was increased to 8.6~nm and 97~GHz of continuous tuning with an intrinsic linewidth of 198~kHz. Even lower linewidths could be achieved by further increasing the cavity length, while the wavelength coverage and tunabilty could be increased by using amplifiers with a wider gain spectrum and by using longer heaters elements.

The available wavelength range of such chip-integrated laser sources can be extended further within the visible using silicon nitride \cite{corato2022widely, siddharth2022near} circuits, while exploring other material platforms such as Al\textsubscript{2}O\textsubscript{3} could enable chip-integrated UV lasers \cite{franken2023hybrid}.

Integrated widely tunable lasers will contribute to integration and upscaling of large and complex optical systems. For example, quantum computer systems \cite{PRXQuantum.2.020343} which today rely on bulky and complex laser systems could benefit from integrated laser sources, while integration into quantum repeater nodes could contribute to future quantum networks \cite{pompili}. Likewise, chip-integrated components such as quantum emitters \cite{schrinner2020integration}, Brillouin lasers \cite{chauhan2021visible} and ion traps \cite{mehta2020integrated} could be combined with on-chip sources.

\section*{Funding}
Rijksdienst voor Ondernemend Nederland (PPS\_2020\_90)

\bibliographystyle{opticajnl}
\bibliography{sample}

\begin{thebibliography}{10}
\newcommand{\enquote}[1]{``#1''}

\bibitem{slussarenko2019photonic}
S.~Slussarenko and G.~J. Pryde, \enquote{Photonic quantum information processing: A concise review,} {\protect\JournalTitle{Applied Physics Reviews}} \textbf{6}, 041303 (2019).

\bibitem{sibson2017chip}
P.~Sibson, C.~Erven, M.~Godfrey, S.~Miki, T.~Yamashita, M.~Fujiwara, M.~Sasaki, H.~Terai, M.~G. Tanner, C.~M. Natarajan, R.~H. Hadfield, J.~L. O’Brien, and M.~G. Thompson, \enquote{Chip-based quantum key distribution,} {\protect\JournalTitle{Nature Communications}} \textbf{8}, 13984 (2017).

\bibitem{degen2017quantum}
C.~L. Degen, F.~Reinhard, and P.~Cappellaro, \enquote{Quantum sensing,} {\protect\JournalTitle{Reviews of Modern Physics}} \textbf{89}, 035002 (2017).

\bibitem{moody20222022}
G.~Moody, V.~J. Sorger, D.~J. Blumenthal, P.~W. Juodawlkis, W.~Loh, C.~Sorace-Agaskar, A.~E. Jones, K.~C. Balram, J.~C. Matthews, A.~Laing \emph{et~al.}, \enquote{2022 {R}oadmap on integrated quantum photonics,} {\protect\JournalTitle{Journal of Physics: Photonics}} \textbf{4}, 012501 (2022).

\bibitem{Taballione2023modeuniversal}
C.~Taballione, M.~C. Anguita, M.~de~Goede, P.~Venderbosch, B.~Kassenberg, H.~Snijders, N.~Kannan, W.~L. Vleeshouwers, D.~Smith, J.~P. Epping, R.~van~der Meer, P.~W.~H. Pinkse, H.~van~den Vlekkert, and J.~J. Renema, \enquote{20-{M}ode {U}niversal {Q}uantum {P}hotonic {P}rocessor,} {\protect\JournalTitle{{Quantum}}} \textbf{7}, 1071 (2023).

\bibitem{somhorst2021quantum}
F.~H.~B. Somhorst, R.~van~der Meer, M.~C. Anguita, R.~Schadow, H.~J. Snijders, M.~de~Goede, B.~Kassenberg, P.~Venderbosch, C.~Taballione, J.~P. Epping, H.~H. van~den Vlekkert, J.~Timmerhuis, J.~F.~F. Bulmer, J.~Lugani, I.~A. Walmsley, P.~W.~H. Pinkse, J.~Eisert, N.~Walk, and J.~J. Renema, \enquote{Quantum simulation of thermodynamics in an integrated quantum photonic processor,} {\protect\JournalTitle{arXiv preprint arXiv:2201.00049}}  (2021).

\bibitem{mahmudlu2023fully}
H.~Mahmudlu, R.~Johanning, A.~Van~Rees, A.~Khodadad~Kashi, J.~P. Epping, R.~Haldar, K.-J. Boller, and M.~Kues, \enquote{Fully on-chip photonic turnkey quantum source for entangled qubit/qudit state generation,} {\protect\JournalTitle{Nature Photonics}} \textbf{17}, 518--524 (2023).

\bibitem{pogorelov2021compact}
I.~Pogorelov, T.~Feldker, C.~D. Marciniak, L.~Postler, G.~Jacob, O.~Krieglsteiner, V.~Podlesnic, M.~Meth, V.~Negnevitsky, M.~Stadler, B.~H\"ofer, C.~W\"achter, K.~Lakhmanskiy, R.~Blatt, P.~Schindler, and T.~Monz, \enquote{Compact ion-trap quantum computing demonstrator,} {\protect\JournalTitle{PRX Quantum}} \textbf{2}, 020343 (2021).

\bibitem{shields2015efficient}
B.~J. Shields, Q.~P. Unterreithmeier, N.~P. de~Leon, H.~Park, and M.~D. Lukin, \enquote{Efficient readout of a single spin state in diamond via spin-to-charge conversion,} {\protect\JournalTitle{Physical Review Letters}} \textbf{114}, 136402 (2015).

\bibitem{roeloffzen2018low}
C.~G.~H. Roeloffzen, M.~Hoekman, E.~J. Klein, L.~S. Wevers, R.~B. Timens, D.~Marchenko, D.~Geskus, R.~Dekker, A.~Alippi, R.~Grootjans \emph{et~al.}, \enquote{Low-loss {Si\textsubscript{3}N\textsubscript{4}} {T}ri{P}lex optical waveguides: {T}echnology and applications overview,} {\protect\JournalTitle{IEEE Journal of Selected Topics in Quantum Electronics}} \textbf{24}, 4400321 (2018).

\bibitem{franken2021hybrid}
C.~A.~A. Franken, A.~van Rees, L.~V. Winkler, Y.~Fan, D.~Geskus, R.~Dekker, D.~H. Geuzebroek, C.~Fallnich, P.~J.~M. van~der Slot, and K.-J. Boller, \enquote{Hybrid-integrated diode laser in the visible spectral range,} {\protect\JournalTitle{Optics Letters}} \textbf{46}, 4904--4907 (2021).

\bibitem{jin2021hertz}
W.~Jin, Q.-F. Yang, L.~Chang, B.~Shen, H.~Wang, M.~A. Leal, L.~Wu, M.~Gao, A.~Feshali, M.~Paniccia, K.~J. Vahala, and J.~E. Bowers, \enquote{Hertz-linewidth semiconductor lasers using {CMOS-ready} ultra-high-{Q} microresonators,} {\protect\JournalTitle{Nature Photonics}} \textbf{15}, 346--353 (2021).

\bibitem{corato2022widely}
M.~Corato-Zanarella, A.~Gil-Molina, X.~Ji, M.~C. Shin, A.~Mohanty, and M.~Lipson, \enquote{Widely tunable and narrow-linewidth chip-scale lasers from near-ultraviolet to near-infrared wavelengths,} {\protect\JournalTitle{Nature Photonics}} pp. 1--8 (2022).

\bibitem{siddharth2022near}
A.~Siddharth, T.~Wunderer, G.~Lihachev, A.~S. Voloshin, C.~Haller, R.~N. Wang, M.~Teepe, Z.~Yang, J.~Liu, J.~Riemensberger, N.~Grandjean, N.~Johnson, and T.~J. Kippenberg, \enquote{Near ultraviolet photonic integrated lasers based on silicon nitride,} {\protect\JournalTitle{APL Photonics}} \textbf{7}, 046108 (2022).

\bibitem{lockingAlbertLisa}
A.~van Rees, L.~V. Winkler, P.~Brochard, D.~Geskus, P.~J.~M. van~der Slot, C.~Nölleke, and K.-J. Boller, \enquote{Long-term absolute frequency stabilization of a hybrid-integrated {I}n{P}-{S}i$_3${N}$_4$ diode laser,} {\protect\JournalTitle{IEEE Photonics Journal}} \textbf{15}, 1502408 (2023).

\bibitem{fan2020hybrid}
Y.~Fan, A.~van Rees, P.~J. Van~der Slot, J.~Mak, R.~M. Oldenbeuving, M.~Hoekman, D.~Geskus, C.~G.~H. Roeloffzen, and K.-J. Boller, \enquote{Hybrid integrated {I}n{P}-{S}i$_3${N}$_4$ diode laser with a 40-{H}z intrinsic linewidth,} {\protect\JournalTitle{Optics Express}} \textbf{28}, 21713--21728 (2020).

\bibitem{kondratiev2017self}
N.~Kondratiev, V.~Lobanov, A.~Cherenkov, A.~Voloshin, N.~Pavlov, S.~Koptyaev, and M.~Gorodetsky, \enquote{Self-injection locking of a laser diode to a high-q wgm microresonator,} {\protect\JournalTitle{Optics Express}} \textbf{25}, 28167--28178 (2017).

\bibitem{guo2022chip}
J.~Guo, C.~A. McLemore, C.~Xiang, D.~Lee, L.~Wu, W.~Jin, M.~Kelleher, N.~Jin, D.~Mason, L.~Chang, A.~Feshali, M.~Paniccia, P.~T. Rakich, K.~J. Vahala, S.~A. Diddams, F.~Quinlan, and J.~E. Bowers, \enquote{Chip-based laser with 1-hertz integrated linewidth,} {\protect\JournalTitle{Science Advances}} \textbf{8}, eabp9006 (2022).

\bibitem{snigirev2023ultrafast}
V.~Snigirev, A.~Riedhauser, G.~Lihachev, M.~Churaev, J.~Riemensberger, R.~N. Wang, A.~Siddharth, G.~Huang, C.~M{\"o}hl, Y.~Popoff, U.~Drechsler, D.~Caimi, S.~H{\"o}nl, J.~Liu, P.~Seidler, and T.~J. Kippenberg, \enquote{Ultrafast tunable lasers using lithium niobate integrated photonics,} {\protect\JournalTitle{Nature}} \textbf{615}, 411--417 (2023).

\bibitem{vanRees:20}
A.~van Rees, Y.~Fan, D.~Geskus, E.~J. Klein, R.~M. Oldenbeuving, P.~J.~M. van~der Slot, and K.-J. Boller, \enquote{Ring resonator enhanced mode-hop-free wavelength tuning of an integrated extended-cavity laser,} {\protect\JournalTitle{Opt. Express}} \textbf{28}, 5669--5683 (2020).

\bibitem{wandt1996external}
D.~Wandt, M.~Laschek, K.~Przyklenk, A.~T{\"u}nnermann, and H.~Welling, \enquote{External cavity laser diode with 40 nm continuous tuning range around 825 nm,} {\protect\JournalTitle{Optics Communications}} \textbf{130}, 81--84 (1996).

\bibitem{sinclair2020silicon}
M.~Sinclair, \enquote{Silicon nitride waveguides and micro-ring resonators for photonic integrated circuits,} Ph.D. thesis, University of Glasgow (2020).

\bibitem{chauhan2021visible}
N.~Chauhan, A.~Isichenko, K.~Liu, J.~Wang, Q.~Zhao, R.~O. Behunin, P.~T. Rakich, A.~M. Jayich, C.~Fertig, C.~W. Hoyt, and D.~J. Blumenthal, \enquote{Visible light photonic integrated brillouin laser,} {\protect\JournalTitle{Nature Communications}} \textbf{12}, 4685 (2021).

\bibitem{Bauters:11}
J.~F. Bauters, M.~J.~R. Heck, D.~D. John, J.~S. Barton, C.~M. Bruinink, A.~Leinse, R.~G. Heideman, D.~J. Blumenthal, and J.~E. Bowers, \enquote{Planar waveguides with less than 0.1 d{B}/m propagation loss fabricated with wafer bonding,} {\protect\JournalTitle{Opt. Express}} \textbf{19}, 24090--24101 (2011).

\bibitem{bogaerts2012silicon}
W.~Bogaerts, P.~De~Heyn, T.~Van~Vaerenbergh, K.~De~Vos, S.~Kumar~Selvaraja, T.~Claes, P.~Dumon, P.~Bienstman, D.~Van~Thourhout, and R.~Baets, \enquote{Silicon microring resonators,} {\protect\JournalTitle{Laser \& Photonics Reviews}} \textbf{6}, 47--73 (2012).

\bibitem{photonics7010004}
K.-J. Boller, A.~van Rees, Y.~Fan, J.~Mak, R.~E.~M. Lammerink, C.~A.~A. Franken, P.~J.~M. van~der Slot, D.~A.~I. Marpaung, C.~Fallnich, J.~P. Epping, R.~M. Oldenbeuving, D.~Geskus, R.~Dekker, I.~Visscher, R.~Grootjans, C.~G.~H. Roeloffzen, M.~Hoekman, E.~J. Klein, A.~Leinse, and R.~G. Heideman, \enquote{Hybrid integrated semiconductor lasers with silicon nitride feedback circuits,} {\protect\JournalTitle{Photonics}} \textbf{7}, 4 (2020).

\bibitem{li2016backscattering}
A.~Li, T.~Van~Vaerenbergh, P.~De~Heyn, P.~Bienstman, and W.~Bogaerts, \enquote{Backscattering in silicon microring resonators: a quantitative analysis,} {\protect\JournalTitle{Laser \& Photonics Reviews}} \textbf{10}, 420--431 (2016).

\bibitem{yariv2002critical}
A.~Yariv, \enquote{Critical coupling and its control in optical waveguide-ring resonator systems,} {\protect\JournalTitle{IEEE Photonics Technology Letters}} \textbf{14}, 483--485 (2002).

\bibitem{Schiemangk:14}
M.~Schiemangk, S.~Spie{\ss}berger, A.~Wicht, G.~Erbert, G.~Tr\"{a}nkle, and A.~Peters, \enquote{Accurate frequency noise measurement of free-running lasers,} {\protect\JournalTitle{Appl. Opt.}} \textbf{53}, 7138--7143 (2014).

\bibitem{Tsuchida:17}
H.~Tsuchida, \enquote{Waveform measurement technique for phase/frequency-modulated lights based on self-heterodyne interferometry,} {\protect\JournalTitle{Opt. Express}} \textbf{25}, 4793--4799 (2017).

\bibitem{schawlow1958infrared}
A.~L. Schawlow and C.~H. Townes, \enquote{Infrared and optical masers,} {\protect\JournalTitle{Physical Review}} \textbf{112}, 1940--1949 (1958).

\bibitem{pintus2023demonstration}
P.~Pintus, J.~Guo, M.~A. Tran, W.~Jin, J.~Liang, J.~Peters, C.~Xiang, O.~J. Ohanian~III, and J.~E. Bowers, \enquote{Demonstration of large mode-hop-free tuning in narrow-linewidth heterogeneous integrated laser,} {\protect\JournalTitle{Journal of Lightwave Technology}} \textbf{41}, 6723--6734 (2023).

\bibitem{franken2023hybrid}
C.~A.~A. Franken, W.~A. P.~M. Hendriks, L.~V. Winkler, M.~Dijkstra, A.~R. do~Nascimento~Jr., A.~van Rees, M.~R.~S. Mardani, R.~Dekker, J.~van Kerkhof, P.~J.~M. van~der Slot, S.~M. García-Blanco, and K.~J. Boller, \enquote{Hybrid integrated near {UV} lasers using the deep-{UV} {A}l$_2${O}$_3$ platform,} {\protect\JournalTitle{arXiv: 2302.11492 [physics.optics]}}  (2023).

\bibitem{PRXQuantum.2.020343}
I.~Pogorelov, T.~Feldker, C.~D. Marciniak, L.~Postler, G.~Jacob, O.~Krieglsteiner, V.~Podlesnic, M.~Meth, V.~Negnevitsky, M.~Stadler, B.~H\"ofer, C.~W\"achter, K.~Lakhmanskiy, R.~Blatt, P.~Schindler, and T.~Monz, \enquote{Compact ion-trap quantum computing demonstrator,} {\protect\JournalTitle{PRX Quantum}} \textbf{2}, 020343 (2021).

\bibitem{pompili}
M.~Pompili, S.~L.~N. Hermans, S.~Baier, H.~K.~C. Beukers, P.~C. Humphreys, R.~N. Schouten, R.~F.~L. Vermeulen, M.~J. Tiggelman, L.~dos Santos~Martins, B.~Dirkse, S.~Wehner, and R.~Hanson, \enquote{Realization of a multinode quantum network of remote solid-state qubits,} {\protect\JournalTitle{Science}} \textbf{372}, 259--264 (2021).

\bibitem{schrinner2020integration}
P.~P.~J. Schrinner, J.~Olthaus, D.~E. Reiter, and C.~Schuck, \enquote{Integration of diamond-based quantum emitters with nanophotonic circuits,} {\protect\JournalTitle{Nano Letters}} \textbf{20}, 8170--8177 (2020).

\bibitem{mehta2020integrated}
K.~K. Mehta, C.~Zhang, M.~Malinowski, T.-L. Nguyen, M.~Stadler, and J.~P. Home, \enquote{Integrated optical multi-ion quantum logic,} {\protect\JournalTitle{Nature}} \textbf{586}, 533--537 (2020).

\end{thebibliography}

\end{document}